\documentclass{jfm}
\usepackage[utf8]{inputenc}
\usepackage{graphicx}
\graphicspath{ {Figures/} }
\usepackage[numbers]{natbib}
\usepackage{amsmath,amssymb,amsfonts,amssymb}
\usepackage{subcaption}
\usepackage{setspace}
\usepackage{float}
\usepackage{microtype}
\usepackage[normalem]{ulem}
\usepackage{ragged2e}
\usepackage{xcolor}
\usepackage{tabularx}
\usepackage{comment}

\DeclareCaptionJustification{justified}{\justifying}
\usepackage{cancel}

\usepackage{hyperref}
\hypersetup{
	hyperindex,
	breaklinks,
	colorlinks=true,
	linkcolor=blue,
	citecolor=magenta,
	bookmarks=true,
	bookmarksopen=true,
	bookmarksopenlevel=2,
	pdfstartpage={1},
	pdfstartview={FitH},
	pdfview={FitH 0},
	pdfauthor={M.A. Nikolaidis, P. J. Ioannou and B. F. Farrell},
	pdftitle={}}

\usepackage{ifthen}

\def\U{\bm{\mathsf{U}}}
\def\Uv{\mathbf{U}}
\def\uv{\mathbf{u}}

\def\C{\bm{\mathsf{C}}}

\def\P{{\bf P}}

\def\Q{\mathbf{Q}}

\def\U{\bm{\mathsf{U}}}

\def\Uv{\boldsymbol{U}}

\def\C{{\bf C}}

\def\P{{\bf P}}

\definecolor{fgreen}{rgb}{0.0, 0.5, 0.0}
\definecolor{dblue}{rgb}{0.2, 0.2, 0.6}
\definecolor{springgreen}{rgb}{0.09, 0.45, 0.27}
\definecolor{dartmouthgreen}{rgb}{0.05, 0.5, 0.06}
\definecolor{egyptianblue}{rgb}{0.06, 0.2, 0.65}
\definecolor{fireenginered}{rgb}{0.81, 0.09, 0.13}
\definecolor{forestgreen}{rgb}{0.0, 0.27, 0.13}
\definecolor{harvardcrimson}{rgb}{0.79, 0.0, 0.09}
\definecolor{amaranth}{rgb}{0.9, 0.17, 0.31}

\newcommand\redsout{\bgroup\markoverwith{\textcolor{red}{\rule[0.5ex]{2pt}{0.4pt}}}\ULon}

\newcommand{\be}{\begin{equation}}
\newcommand{\ee}{\end{equation}}
\newcommand{\bdm}{\begin{equation*}}
\newcommand{\edm}{\end{equation*}}
\newcommand{\bea}{\begin{eqnarray}}
\newcommand{\eea}{\end{eqnarray}}

\newcommand{\partialf}[2]
{
 \ifthenelse{\equal{#1}{}}{\frac{\partial}{\partial #2}}{\frac{\partial #1}{\partial #2}}
}

\newcounter{saveeqn}%

\def\bt{\tilde{\beta}}

\def\st{\sin{\vartheta}}

\def\xv{\mathbf{x}}

\def\Uv{\mathbf{U}}
\def\uv{\mathbf{u}}

\newcommand{\defn}{\ensuremath{\stackrel{\mathrm{def}}{=}}}
\renewcommand{\equiv}{\defn}

\renewcommand{\U}{\mathbf{U}}
\renewcommand{\u}{\mathbf{u}}

\shorttitle{SSD of WWC turbulence}
\shortauthor{Farrell \& Ioannou}

\title{Statistical state dynamics modes and equilibria underlie the structure and mechanism of  wide channel Couette turbulence}

\author{Brian F.~Farrell\aff{2},
 Petros J. Ioannou\aff{1,2}
\corresp{\email{pjioannou@phys.uoa.gr}}}
\affiliation{\aff{1}Department of Physics, National and Kapodistrian University of Athens, Athens, Greece
\aff{2}Department of Earth and Planetary Sciences, Harvard University, Cambridge, U.S.A. 
}

\begin{document}

\maketitle

\begin{abstract}
Wide channel Couette (WCC) turbulence is striking in being dominated by a large-scale spanwise periodic structure
composed of streamwise streaks and associated roll superstructures. This apparent equilibrium is shown
in this work to correspond to a fixed point solution of the Navier-Stokes equations expressed in the statistical state 
dynamics (SSD) framework.  Moreover, this fixed point solution is found to be 
rank-three, consisting of one analytically identified roll-streak structure (RSS)
constituting the first cumulant of the SSD, and two analytically determined eigenmodes supporting the 
second cumulant. This minimal representation captures both the structure and the dynamics of WCC turbulence, 
while the remaining spectral components contribute negligibly to the equilibrium dynamics.  Turbulent Couette flows
other than WCC can be understood to be limit cycle and chaotic extensions supported by the same underlying mechanism as
the fixed point WCC turbulence except for the two eigenmodes supporting WCC turbulence being replaced by two Floquet modes 
and two Lyapunov vectors, respectively.  These results provide an analytic solution for turbulence in Couette flow.

\end{abstract}
\section{Introduction}

It is advantageous in constructing a theory for a physical phenomenon to begin with the 
simplest manifestation of the phenomenon and to build understanding of its more 
complex manifestations
on the foundation of a comprehensive understanding of 
the simplest example. In the case of wall turbulence, the simplest canonical example is 
the turbulence that develops in wide channel Couette flow (WCC).

Experimental observations
\cite{Bech-1994,Papavasiliou-1997,Tillmark-1995,Tillmark-1998,Kitoh-2005,Kitoh-2008}
and DNS simulations of WCC at a wide range of Reynolds numbers 
\cite{Komminaho-etal-1996,Tsukahara-etal-2006,Pirozzoli-etal-2014,Avsarkisov-etal-2014, Lee-Moser-2018,Hoyas-Oberlack-2024}
have revealed that Couette turbulence is dominated by an apparent equilibrium state consisting of large scale roll streak structures (RSS).
These structures represent a fundamental organization of the turbulent flow field, with the streaks exhibiting 
characteristic spanwise wavelengths and maintaining coherence over extended streamwise distances.
As the channel becomes wider in the spanwise direction so that the dynamics becomes increasingly independent of 
spanwise boundary effects, the temporal variability of the RSS diminishes significantly \cite{Pirozzoli-etal-2014}. 
This reduction in time-dependence strongly suggests that the flow approaches an equilibrium state in which 
the large-scale RSS organization achieves a time independent balance among its energy production, transport, 
and dissipation processes. 
The emergence of this fixed equilibrium turbulent state in WCC provides a unique opportunity to analyze
the  fundamental structures and dynamics of wall turbulence in its simplest and most transparent manifestation.

These equilibrium structures in WCC are not equilibria of the Navier-Stokes equations (NSE) formulated in velocity 
state variables, which is the reason that, despite the striking evidence for a fixed point structure presented by  
DNS of WCC, these equilibria have remained unexplained.  The reason the analytic structure of these 
equilibria have remained unidentified is that the instabilities underlying these structures are nonlinear. 
Identification of the  structure underlying turbulence in shear flow and the dynamics of this structure, 
which supports the turbulent state,  follows from formulating the NSE so that this  
structure can be identified using linear perturbation theory.  In order to identify the turbulent state  it is also required that the nonlinear equilibration of this 
instability be followed analytically to its equilibration as a turbulent state.  
The pivotal first step required to execute this program is to formulate the NSE in nonlinear variables so that
the nonlinear instability underlying the turbulence can be identified using linear perturbation theory.  It is 
then straightforward to use these same equations to find the associated equilibria proceeding from these instabilities which
can be identified as the analytic expression of the turbulent state.
The formulation of the NSE in nonlinear variables that allows this analysis is the statistical state dynamics (SSD) formulation and specifically closure of the NSE in a cumulant expansion at second order.  The stochastic structural stability theory (S3T) formulation is the closure of the NSE SSD at second order in a cumulant expansion in which the averaging operator 
is chosen to be the streamwise average and the fluctuation-fluctuation interactions are parameterized as a white in time stochastic process.
The  S3T formulation  
was originally developed as a component of a theory for  the spontaneous emergence of jets in turbulent planetary atmospheres
and  bears the legacy acronym, S3T,  of that use \cite{Farrell-Ioannou-2003-structural}.
S3T  incorporates the fundamental dynamics of wall-turbulence while retaining maximal analytic clarity and simplicity.
In this closure the covariance of the fluctuations appears as a variable.  Linear perturbation of the covariance as a variable allows identification of the nonlinear instability underlying the turbulence.
  The modes identified by stability analysis of  S3T equilibria have previously been shown to predict the perturbation  RSS that arises in 
transitional flows in the presence of free-stream turbulence. Study of the mechanism sustaining this
perturbation RSS revealed that it is also the mechanism underlying the
self sustaining process (SSP) in fully turbulent flows \cite{Farrell-Ioannou-2012,Farrell-Ioannou-2017-bifur, Nikolaidis-etal-2024}.   


%
Although the S3T SSD is the most simple non-trivial closure of the NSE SSD, study of SSD dynamics even in S3T remains numerically challenging due to the high dimensionality of
the S3T state space. In the  S3T formulation of the SSD dynamics of the NSE in wall-turbulence, 
it is necessary to advance both the first cumulant, which has
dimension $O(N_y \times N_z)$, where  $N_y$ and $N_z$ denote the number of spectral 
components required
to resolve  the state 
in the wall-normal   and  spanwise directions, respectively, 
and  the second cumulant which, in its simplest manifestation, has dimension
$O(N_y^2 \times N_z^2)$. 
This quadratic scaling of the second cumulant makes direct numerical integration computationally challenging.
We find that for most purposes the infinite ensemble 
required by S3T dynamics can be adequately approximated by a finite ensemble
\cite{Farrell-Ioannou-2017-bifur}. 
Moreover, we show below that the ensemble representation of the S3T dynamics without explicit stochastic closure 
not only 
reproduces the infinite-ensemble dynamics of S3T exactly, but also reveals structural information 
about the second cumulant that remains obscured in the standard cumulant 
representation. Specifically, the ensemble formulation exposes the rank structure 
and modal composition of the second cumulant, providing physical insight into the mechanisms 
sustaining the turbulent equilibria. 

In the following  section
we derive the ensemble representation of NSE SSD dynamics and explain
how finite ensembles can capture the infinite-ensemble limit of S3T dynamics. 
In the remaining sections, we apply this framework to  WCC turbulence.

\section{Ensemble formulation of the statistical state dynamics of plane Couette turbulence}

Consider an incompressible plane Couette flow of density $\rho$
in a  channel periodic in the streamwise, $x$, and spanwise, $z$, direction with boundary conditions  $\u(x+L_x,y,z)=\u(x,y,z)$, $\u(x,y,z+L_z)=\u(x,y,z)$
where $L_x$ and $L_z$ are the  non-dimensional  channel extents  in $x$ and  $z$ and with cross-stream
channel walls at $y/h=\pm1$, with $\pm U_w \hat{\bf x}$ the velocity  at $y/h=\pm 1$ respectively where  ($\hat{\xv}$ is the unit vector in the $x$-direction). The Reynolds number is $R= U_w h/\nu$, with $\nu$ the kinematic viscosity.


The velocity field $\uv$ is non-dimensionalized by $U_w$, the pressure field by $\rho U_w^2$, the spatial variables $x$, $y$, $z$,  by $h$ and 
time by $h/U_w$.
With this choice of non-dimensional variables the NSE governing the flow  and without any change of notation becomes:
\begin{equation}
\label{eq:NS0}
\partial_t\uv  + \uv \cdot \nabla \uv  + \nabla p -  \Delta \uv/R =0 ~~,~~~\nabla \cdot \uv=0~.
\end{equation}

Denoting  an averaging operator satisfying Reynolds conditions \citep{Monin-Yaglom-1971} as $\{\cdot\}$,  the velocity and pressure fields in \eqref{eq:NS0} can be partitioned into mean, denoted by $\Uv(y,z,t)=\{\u\}$, $P(y,z,t)=\{p\}$
and  deviations $\uv'(x,y,z,t)$, $p'(x,y,z,t)$ from this mean (referred to as fluctuations)
so that  $\uv = \Uv + \uv'$ and $p= P(y,z,t)+ p'(x,y,z,t)$.  
Partition of the state into mean and fluctuations allows the  NSE  to be expressed as separate equations for the   evolution of the mean and the fluctuations:
\begin{subequations}
\label{eq:NS}
\begin{align}
&&\partial_t\boldsymbol{U}  + \boldsymbol{U} \cdot \nabla \boldsymbol{U}   + \nabla P -  \Delta \boldsymbol{U}/R  =- 
\{ \boldsymbol{u}' \cdot \nabla \boldsymbol{u}' \}\ ,
\label{eq:NSm}\\
&& \partial_t\boldsymbol{u}'+   \boldsymbol{U} \cdot \nabla \boldsymbol{u}' +
\boldsymbol{u}' \cdot \nabla \boldsymbol{U}  + \nabla p' -  \Delta  \boldsymbol{u}'/R = - (  \boldsymbol{u}' \cdot \nabla \boldsymbol{u}' - 
\{\boldsymbol{u}' \cdot \nabla \boldsymbol{u}'\} \,)~,
 \label{eq:NSp}\\
&& \nabla \cdot \boldsymbol{U} = 0\ ,\ \ \ \nabla \cdot \boldsymbol{u}' = 0\ , \label{eq:NSdiv0}
\end{align}\label{eq:NSE0}\end{subequations}
with corresponding boundary conditions. 
The averaging operators we will consider are spatial averages over coordinates and ensemble averages. 
Spatial averages will be denoted by square brackets with a subscript indicating the independent variable over which the average is taken; for example, streamwise averages are defined as 
$[\,{\cdot}\,]_x=L_x^{-1} \int_0^{L_x} {\cdot}\  dx$, 
while ensemble averages over independent realizations of the flow field will be denoted by angle brackets $\langle \cdot \rangle$.

In formulating the SSD for our study we adopt for Reynolds averaged  operator the streamwise mean and interpret \eqref{eq:NSm} as the first cumulant
with that Reynolds averaged operator and consider that an ensemble of fluctuations that evolve in the mean field $(\U,P)$ according to \eqref{eq:NSp} interact with the mean through the Reynolds stress divergence produced by their ensemble average.
In order to produce an SSD closed at second order it is necessary to
parameterize the contribution of the fluctuation-fluctuation interactions $(  \boldsymbol{u}' \cdot \nabla \boldsymbol{u}' - 
\{\boldsymbol{u}' \cdot \nabla \boldsymbol{u}'\} \,)$ in the ensemble fluctuation dynamics in a state independent manner.  We have in the past used a stochastic parameterization for this term that excites the fluctuation equation with independent temporally white structures with spatial correlation $\Q$.  Although we make use of results obtained using this parameterization, 
in this work we primarily ignore this term by setting $\Q=0$. The $N$-ensemble 
SSD is thus governed by the equations
\begin{subequations}
\label{eq:eRNL}
\begin{align}
&\partial_t\boldsymbol{U}_N+ \boldsymbol{U}_N \cdot \nabla \boldsymbol{U}_N   + \nabla P -  \Delta \boldsymbol{U}_N/R = 
- {\Large \langle} \left [\boldsymbol{u}_n' \cdot \nabla \boldsymbol{u}_n'\right ]_x  {\Large \rangle}_N 
\label{eq:eRNLm}\\
 &\partial_t\boldsymbol{u}_n'+   \boldsymbol{U}_N \cdot \nabla \boldsymbol{u}_n' +
\boldsymbol{u}_n' \cdot \nabla \boldsymbol{U}_N  + \nabla p_n' -  \Delta  \boldsymbol{u}_n'/R
= 0\ ,~~n=1,2,\cdots,N~, \label{eq:eRNLp}\\
& \nabla \cdot \boldsymbol{U}_N = 0\ ,\ \ \ \nabla \cdot \boldsymbol{u}_n' = 0\,,~~n=1,2,\cdots,N~,\label{eq:eRNLdiv0}
\end{align}
\end{subequations}
where $n=1,\cdots,N$ indicates the ensemble member while $\U_N$ is the mean flow shared in common by the $N$ fluctuation ensemble members
 and $\left < \cdot \right >_N $ indicates an average over the $N$ ensemble members. In the limit $N \to \infty$ we obtain a
SSD closed at second order referred to as the S3T system with $\Q=0$, which has as its variables
the first and second cumulants of the velocity fields, $(\U,\C)$. 
We refer to system \eqref{eq:eRNL} as the restricted nonlinear system (RNL) with $N$ ensemble members and denote it ${RNL}_N$.



By choosing the streamwise mean 
as the averaging operator we implicitly make 
the assumption that the statistical state is statistically homogeneous in the streamwise direction and that this statistical state is stable.  The
fluctuation fields  can then be expanded in a Fourier series:
\begin{equation}
\uv' (x,y,z,t)= \sum_{n \in \mathbb{Z}_{\ne 0}}   \hat{\uv}'_{\alpha_n} (y,z,t)\,e^{i \alpha_n x} \ ,
\label{eq:fourier}
\end{equation}
with $\alpha_n=n \alpha =2 \pi n/ L_x$ the streamwise wavenumber,
and correspondingly the second cumulant of the S3T 
streamwise statistically homogeneous  state between point
$1\equiv(x_1,y_1,z_1)$ and point $2\equiv (x_2,y_2,z_2)$,
as 
\begin{gather*}
C_{ij} (1,2,t)= \langle u_i'(x_1,y_1,z_1,t) u_j'(x_2,y_2,z_2,t) \rangle 
= \sum_{n=-\infty}^\infty C_{ij,n}(1,2,t)e^{i \alpha_n(x_1-x_2)},
\end{gather*}
with
\begin{equation}
C_{i j,n}(1,2,t)= \left < \hat{u}_{i,\alpha_n}'(y_1,z_1,t) \hat{u}_{j,\alpha_n}'^*(y_2,z_2,t)  \right >~,
\label{eq:covar}
\end{equation}
the same time  covariance of the $n$-th Fourier component of the  velocity fluctuations, $\hat{u}_{i,\alpha_n}'$,
with the indices $i$ and $j$ indicating the velocity  components of   $\uv'$ ($*$ denotes complex conjugation).  
The S3T SSD equations having as variables the first cumulant $\U$ and the second cumulant the Reynolds averaged fluctuation-fluctuation covariance, $\C$, can then be 
written by forming an equation for the evolution of $\C$  from \eqref{eq:eRNLp} and expressing the Reynolds stress forcing in the mean equation \eqref{eq:eRNLm} in terms of $\C$ (cf. \cite{Farrell-Ioannou-2012,Farrell-etal-2016-PTRSA}). The covariance of the RNL$_N$ system is formed by averaging the covariance of the $N$ ensemble members which approximates the infinite ensemble covariance forming the second cumulant of the
S3T SSD state \eqref{eq:covar}:
\begin{gather}
    C_{N, i j,n}(1,2,t)=\frac{1}{N} \sum_{k=1}^N  \hat{u}_{(i,\alpha_n),k}'(y_1,z_1,t) \hat{u}_{(j,\alpha_n),k}'^*(y_2,z_2,t) ~.
    \label{eq:covarN}
    \end{gather}
This covariance is at most rank $N$, and  converges  to the infinite ensemble covariance, $\C$,  of the S3T SSD as $N \to \infty$.  

In our study of the SSD of wall-turbulence
we have  frequently reported results from RNL$_1$ simulations  (cf.  \cite{Thomas-etal-2014,Farrell-etal-2016-VLSM,Farrell-Ioannou-2017-sync}).
The RNL$_1$ system is the approximation to the S3T SSD obtained using
a single-member ensemble of fluctuations.  In the case of \eqref{eq:eRNL}, in which the  fluctuation-fluctuation interactions are absent, 
this results in a fluctuation covariance matrix, $\C$, of rank 1.
Although the term involving fluctuation-fluctuation interactions in \eqref{eq:NSp}
has been set to zero in \eqref{eq:eRNLp}, the quasi-linear RNL$_1$ system sustains a realistic turbulent state (cf. \cite{Bretheim-etal-2015,Farrell-etal-2016-VLSM,Farrell-etal-2016-PTRSA})  
supported by a small set of fluctuations that are spontaneously maintained by the dynamics and that thereby identify the active subspace supporting RNL$_1$ turbulence \citep{Thomas-etal-2015,Nikolaidis-Ioannou-2022}.
%
In a similar manner, the quasi-linear RNL$_N$ and S3T equations maintain a realistic self-sustaining turbulence supported by a small set of fluctuation structures  at a few streamwise wavenumbers \citep{Farrell-Ioannou-2012, Thomas-etal-2015,Farrell-etal-2016-VLSM}. 
  This subspace of fluctuations  sustaining the turbulence is spontaneously identified by the dynamics and 
  corresponds to the active subspace of the turbulent state, a concept that was introduced on heuristic grounds by Townsend \citep{Townsend-1961,Townsend-1976}.  This active subspace  is identified as the subspace spanned by the small set of Lyapunov vectors 
with zero Lyapunov exponent that survive when the fluctuation state is evolved by the fluctuation
equations \eqref{eq:eRNLp} with the first cumulant, $\U(y,z,t)$, supported by the turbulence. This follows directly from the quasilinearity of the S3T/RNL SSD on reflecting that at equilibrium fluctuations
with negative exponent have vanished, while existence of equilibrium requires that no fluctuations with positive exponent remain.
This implies that the first cumulant of the turbulent flow in S3T/RNL SSD turbulence
(with $\Q=0$) has been regulated by the two-way mean-fluctuation interaction to neutrality,
providing proof of the validity of an extended version of the
Malkus conjecture \citep{Malkus-1956,Reynolds-Tiederman-1967} that turbulent mean flows are regulated to a state
of perturbation neutrality.
This neutrality of the first cumulant in RNL is reflected in  the corresponding DNS 
by the first cumulant being slightly unstable  as required to support the energy cascade to higher streamwise
wavenumbers that is present in DNS but is not present in RNL \citep{Nikolaidis-etal-Madrid-2018,Nikolaidis-Ioannou-2022}.

\begin{figure}
    \begin{subfigure}{\textwidth}
            \centering
            \includegraphics[width=0.8\textwidth]{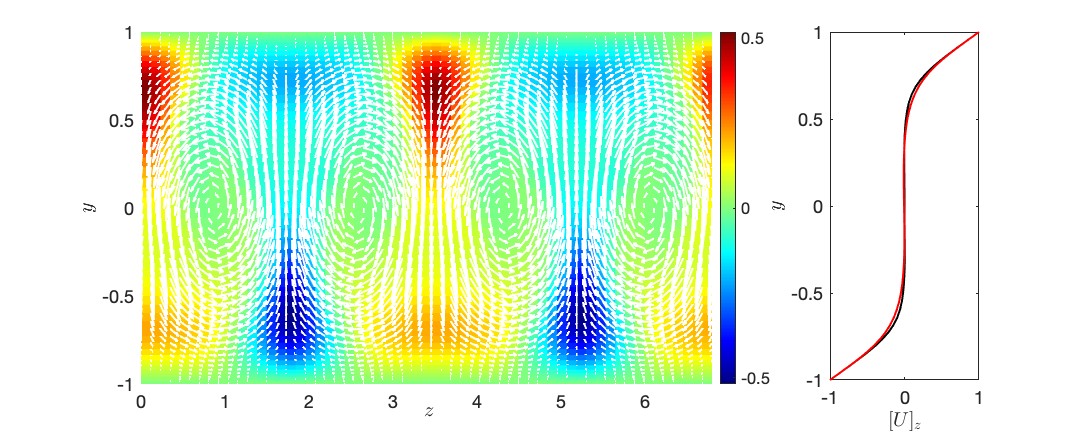}
            \label{fig:P12}
    \end{subfigure}%
      \hfill
    \begin{subfigure}{\textwidth}
            \centering
            \includegraphics[width=0.8\textwidth]{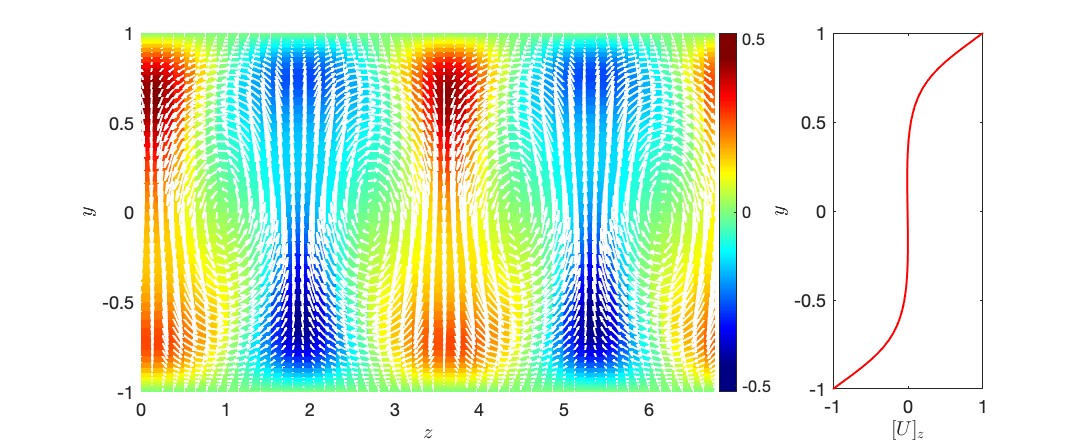}
    \label{fig:P11}            
          \end{subfigure}
    \caption{Top left: Streak component of the streamwise velocity $U_s(y,z)=U(y,z)-[U(y,z)]_z$ (color contours)) and vectors
    of the mean  $(W(y,z),V(y,z))$ velocity of the WCC S3T equilibrium. Top right: The  mean turbulent profile, $[U(y,z)]_z$, of the WCC equilibrium as a function of $y$ (black). The corresponding mean turbulent profile from  DNS is shown in red. The spanwise separation distance between the low-speed streaks is $z=3.5$.   
    Bottom  left: Corresponding  time-mean streak component $[U_s(y,z)]_t$ and vectors
    of the time-mean  $([W(y,z)]_t,[V(y,z)]_t)$  obtained from DNS.  Bottom right: The  time and spanwise averaged streamwise flow, $[U(y,z,t)]_{t,z}$, from DNS. For a channel with  $z=2.2 \pi$ at $R=600$. }
    \label{fig:S3T}
\end{figure}

It follows that the covariances, $\C_N$ and $\C$, associated with RNL$_N$ and S3T respectively, have small rank  corresponding to that of the set of zero Lyapunov exponent fluctuation structures comprising the active subspace. 
Demonstration that a small set of analytically known fluctuations with associated low rank covariance, $\C$, supports realistic shear turbulence in S3T suggests that NSE turbulence is essentially supported on a closely related limited set of identifiable structures, 
however much inessential structures produced as a result of fluctuation-fluctuation nonlinearity may obscure this underlying dynamical simplicity
\citep{Nikolaidis-Ioannou-2022}.
As noted above, the first step in this identification is to realize that the rank of the covariances $\C_N$ (or $\C$) is equal to the number of ``neutral modes'' supported in the second cumulant by the first cumulants $\U_N$ (or $\U$) (zero exponent Lyapunov vectors, zero real part eigenvalue for Floquet vectors or eigenmodes in the cases of chaotic, limit cycle or fixed point turbulence, respectively). 

\begin{figure}
    \centering
    \includegraphics[width=1.0\linewidth]{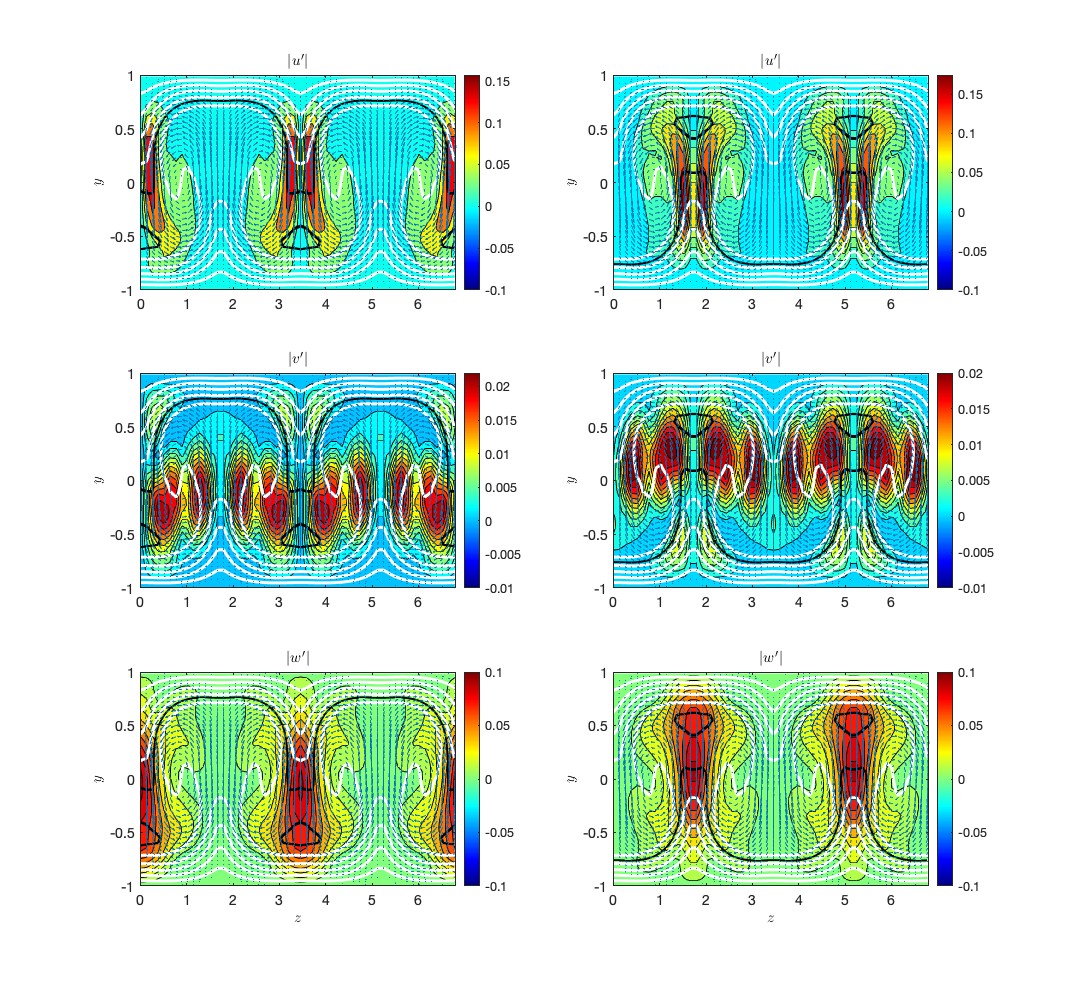} 
    \caption{Velocity field
    of the two neutral modes that support the WCC S3T equilibrium shown in the top panel of Fig. \ref{fig:S3T}. 
    Left panels from top to bottom:  
    contours of the $|u'|$, $|v'|$ and $|w'|$ velocity of the neutral mode with phase speed
    $c_r=0.106$
    at a $y-z$ section of the flow.  The black line indicates the critical layer line for this mode. Also shown in white are contours of $U(y,z)$ and in blue vectors 
    of  $(W,V)$ of the mean flow of the S3T equilibrium. Right panels  from top to bottom:
    the velocity fields of the neutral mode with $c_r=-0.106$ at the corresponding 
    $y-z$ section of the flow.}
    \label{fig:P2}
\end{figure} 

As discussed, the covariance, $\C$, of the S3T
state with $\Q=0$ has  rank equal to the number of Lyapunov vectors with zero Lyapunov exponent when $\C$ is evolved with the 
first cumulant, $U(y, z, t)$, supported by the S3T turbulence,
while the rank of $\C_N$ of the corresponding  RNL$_N$ simulations is naturally
limited by $N$, the number of ensemble members comprising the simulation. This implies that as $N \to \infty$ and the
RNL$_N$ converges to the S3T,   the rank of $\C_N$  becomes  constant and equal to $N_0$,  the rank of the S3T $\C$ , 
assuming, of course, that $\C$ has finite rank, as is the case in all our viscous channel flow simulations. This implies that convergence 
to the infinite ensemble S3T SSD state 
with $\Q=0$  is already obtained by an RNL$_N$ with $N$ 
equal to or greater than the finite rank, $N_0$, of the $\C$ of the S3T.
This result imposes an extraordinary requirement on the structure
of both the first cumulant and the $N_0$ fluctuations comprising the
fluctuations supporting the turbulent state. While this $N_0$ is not initially
known it is easily determined from RNL$_N$ simulations which 
reveal that for some $N$ the rank of $\C_N$
becomes constant and does not increase with increase in the number of ensemble members retained; this constant rank being the
rank, $N_0$, of the infinite ensemble S3T.

In order to appreciate the extraordinary 
requirements that are necessitated by the
convergence of all RNL$_N$ with $N\ge N_0$ to the S3T state 
consider for simplicity the case of fluctuations being supported by a single streamwise wavenumber, $k$, and that the rank  of the S3T covariance, $\C$, is $N_0=2$, so that the fluctuation state  is composed of linear combinations
of the two  normalized\footnote{If the neutral vectors are time-dependent the vectors are normalized at a specified time.}  neutral Lyapunov vectors, with Fourier components
$\hat \u_{1,k}'$ and $\hat \u_{2,k}'$.  For this example, in the $\alpha$-th ensemble member RNL$_N$ with $N \ge N_0$ the velocity components of the fluctuation state are:
\begin{gather}
\lambda_\alpha \hat{u}_{1 i, k}'(y,z,t) + \mu_\alpha \hat{u}_{2 i, k}'(y,z,t)~,
\label{eq:sum}
\end{gather}
and  the $N_0$ rank covariance
of the RNL$_N$ simulation
is 
\begin{eqnarray}
  C_{N, i j,n}(1,2,t) &=&
  \frac{1}{N} \sum_{\alpha=1}^N  
 \left ( \lambda_\alpha \hat{u}_{1 i, k}'(1) + \mu_\alpha \hat{u}_{2 i, k}'(1) \right ) 
 \left ( \lambda_\alpha \hat{u}_{1 j, k}'(2) + \mu_\alpha \hat{u}_{2 j, k}'(2) \right )^*\nonumber \\
& = & \Lambda \hat{u}'_{1 i, k}(1) \hat{u}_{1 j, k}^{'*}(2)
+ M \hat{u}'_{2 i, k}(1) \hat{u}_{2 j, k}^{'*}(2) 
+  \Phi \hat{u}_{1 i, k}^{'}(1)\hat{u}_{2 j, k}^{'*}(2) +\Phi^*  \hat{u}_{1 i, k}^{'*}(2)\hat{u}_{2 j, k}^{'}(1) \nonumber
\end{eqnarray}
with 
\begin{gather*}
\Lambda =
 \frac{1}{N} \sum_{\alpha=1}^{N}|\lambda_\alpha|^2 ~~,
~~M =
 \frac{1}{N} \sum_{\alpha=1}^{N}  | \mu_\alpha|^2~~, \Phi = \frac{1}{N} \sum_{\alpha=1}^{N}  \lambda_\alpha \mu_\alpha^*  .
\end{gather*}
The sum \eqref{eq:sum} of the neutral vectors will be different for each member of the $N$-ensemble (with measure zero exception), but in aggregate, these neutral vector sums must comprise an ensemble supporting for any $N\ge N_0$ and
any possible combination of $\lambda_\alpha$ and $\mu_\alpha$ 
that produces the same $\Lambda$ and  $M$, the same first cumulant by their combined Reynolds stresses, which is determined by the mean fluctuation covariance, $\C_N$.  This requires that the effective feedback regulator operating between the first and second cumulant has been able to adjust the  structure of the Lyapunov vectors 
and their coefficients to satisfy
\begin{gather}
\Phi = \frac{1}{N} \sum_{\alpha=1}^{N}   \lambda_\alpha \mu_\alpha^*  =0 ~,
\label{eq:C}
\end{gather}
 so that the components of the covariance arising from the interference of the modes makes zero contribution to the net Reynolds stress divergence
in \eqref{eq:eRNLm},
i.e. so that the modes do not interfere with each other, and thus produce additional Reynolds stresses at double their frequencies which is incompatible with existence of a fixed point.
This condition corresponds to the requirement 
that the sum of the Reynolds stresses produced by the modes in isolation  supports the first cumulant.
In other words, the
turbulent state  sustained in any $RNL_N$
simulation with $N \ge N_0$ is equivalent to the state
sustained in an $N_0$ ensemble simulation
with fluctuations (in the case of $N_0=2$)  
$\sqrt{\Lambda}~ \hat \u_{1,k}'$ in the first ensemble member and
$\sqrt{M}~ \hat \u_{2,k}'$ in the second ensemble member.
Only when these requirements are met does the RNL$_{N}$ simulation ensemble with any $N \ge N_0$ produce the same first cumulant, $\U$, and consequently converge at $N=N_0$ to the full S3T.  
We will demonstrate in this paper
how these requirements are enforced by the effective feedback regulator operating between the first and second cumulant in the case of an S3T turbulent state that is a stable fixed point.

\begin{figure}
    \centering
    \includegraphics[width=0.8\linewidth]{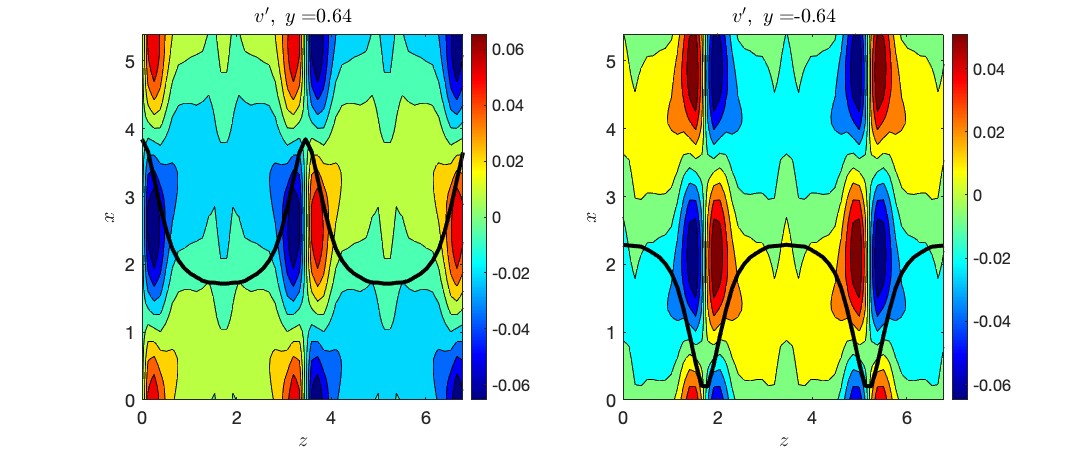}     
    \caption{Cross-stream velocity component, $v'$, of
    the neutral modes shown  in the two columns of Figure \ref{fig:P2}.  The $v'$ component is shown on a $x-z$ cross-section at the level of the respective streak maxima, at $y=\pm 0.64$. The respective streamwise flow at this level is sketched in black. 
    This figure shows that the two modes
    are sinuous oblique waves about their respective streaks. }
    \label{fig:P3}
\end{figure}

As mentioned above, in the S3T state 
the regulator needs to enforce the neutrality of
the first cumulant and also ensure that the sum of Reynolds stresses arising from the neutral vectors in the ensemble support
the first cumulant.   Remarkably, at the same time, the regulator must arrange that these neutral vectors comprise a set with vanishing sum over the cross-covariance between the neutral modes of each of the ensemble members so that there is no fluctuation of the converged first cumulant state due to interference between the supporting neutral vectors that would produce fluctuating Reynolds stresses perturbing the first cumulant structure at the beat frequency of the neutral modes.  We note that this second requirement for vanishing of the sum over the cross terms in the ensemble fluxes is a more difficult task for the regulator to achieve, which confines the fixed-point solutions to a limited region of parameter space.

We conclude that the  $RNL_{N_0}$ ensemble identifies the analytical structure of the full S3T  Couette turbulent state.  This S3T turbulent state is made up of a mean flow and a set of neutral Lyapunov vectors supported by this mean flow.  
The configuration of these elements to form a fixed point, a limit cycle, or chaotic turbulence is determined by the regulator.  It is possible for the regulator to maintain the turbulence in one or another of these turbulent states depending on parameters, initial conditions, or perturbations that are external or internally generated.

We reiterate that by choosing the streamwise averaging operator for our SSD we have assumed that plane Couette turbulence is statistically homogeneous in the streamwise direction but have not enforced statistical homogeneity in the spanwise direction.  This choice may be unexpected given that it has been traditionally assumed that the statistical mean state in plane shear turbulence is both spanwise and streamwise independent, this assumption being in accord with the homogeneity of the problem in both the streamwise and spanwise directions and with the observation that means obtained either as an unconditional ensemble average over flow realizations or as a long time average of a single flow realization lack spanwise  structure.
However, unconditional ensemble averaging of turbulent states is not an appropriate Reynolds average for the formulation of the SSD of shear turbulence because SSD modal instabilities result in spontaneous symmetry breaking leading to emergence of spanwise structure.  
Consistently, it is well known that a turbulent state  is not sustained in  shear flow when the spanwise component of 
structures elongated  in the streamwise direction  is filtered out \citep{Jimenez-Pinelli-1999}. This result reveals that spontaneous symmetry breaking in the spanwise direction is intrinsic to the turbulent state.  Expressing the NSE in cumulant form rather than in traditional velocity state variable form allows identification of these spanwise symmetry-breaking modal structures underlying turbulence in shear flow \cite{Farrell-Ioannou-2012}.

\begin{figure}
    \centering
    \includegraphics[width=1.0\linewidth]{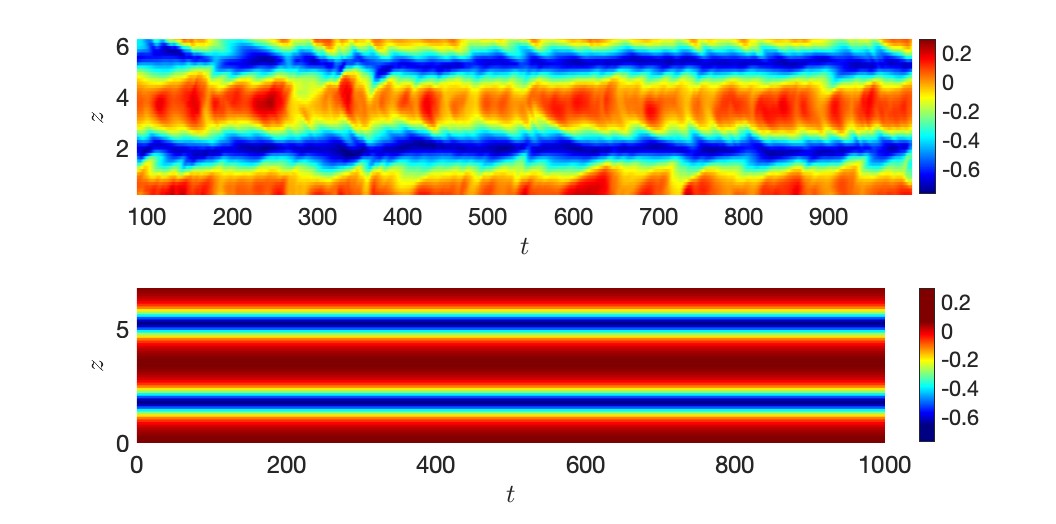}    
 \caption{Top panel: Mean streamwise velocity as a function
    time, $t$, and spanwise distance, $z$, at cross-stream level $y=-0.7$ in a DNS of pCf at $R=600$ 
    with $L_z=2.2 \pi$, $L_x=1.75 \pi$ in which the streamwise mean spanwise velocity, $W$,
    has been filtered out in order to prevent drifting of the streaks. This figure shows the persistence of the streaks in the DNS of this Couette turbulence. 
    Bottom panel: The corresponding S3T equilibrium.}
    \label{fig:P0}
\end{figure}

\section{Methods}

Our SSD study of Couette turbulence  will be at Reynolds number $R=600$ in  periodic channels with streamwise periodicity $L_x=1.75 \pi$
and various spanwise periodicities $L_z$, but principally in a channel with spanwise periodicity $L_z=2.2 \pi$. 
For the numerical integration of the full NSE equations \eqref{eq:NS} and the  RNL$_N$ equations
\eqref{eq:eRNL} the dynamics is expressed in the form of
evolution equations for the wall-normal vorticity and the Laplacian of
the wall-normal velocity, with spatial discretization and Fourier
dealiasing in the  spanwise direction and Chebychev polynomials
in the wall-normal direction as in Kim et al.~\citep{Kim-etal-1987}. Time stepping was
implemented using the third-order semi-implicit Runge-Kutta method.  Equations \eqref{eq:NS} and \eqref{eq:eRNL}
were integrated with the vectorized ensemble direct numerical simulation codes developed by  
M-A. Nikolaidis \citep{Nikolaidis-2024} with a discretization of $N_z=60$  equally-distanced points in $z$ when $L_z \le 2.2 \pi$,  and  proportionally adjusted number of points in $z$ when
$L_z >2.2 \pi$, 
and $N_y=53$ Chebyshev collocation points in $y$.

While NSE turbulence sustains a full spectrum fluctuation field,  the fluctuation field in  RNL$_N$ turbulence in this channel is supported by fluctuations
at  only the fundamental  streamwise wavenumber  $\alpha=2 \pi/L_x$, as this is the streamwise wavenumber of the fluctuations with zero Lyapunov exponent and
therefore the only streamwise Fourier component supporting  
RNL turbulence at this Reynolds number \cite{Farrell-Ioannou-2017-sync,Nikolaidis-Ioannou-2022}.
This single  fluctuation streamwise wavenumber identifies the streamwise component of the active subspace supporting turbulence in the RNL of this flow; 
the remaining streamwise wavenumbers being passive, as revealed by their having negative Lyapunov exponent.
Consistently, for the RNL$_N$ simulations
the code has been modified to retain two harmonics in $x$, $k_x=0$ and $k_x=1 \alpha$.

As the spanwise spatial period, $L_z$, increases, the power of the spanwise-uniform component  of the flow, corresponding to the spanwise
wavenumber $k_z=0$, decreases and approaches zero in the limit of infinite  spanwise width.  
The $k_z=0$ component of the flow arises from all the triad interactions between Fourier modes of the form
$(k_x,k_z) + (k_x',-k_z)$; in  RNL simulations,  these interactions are limited to  two types: $(0,k_z)+(0,-k_z)$ and  $(0,k_z)+(\alpha,-k_z)$.
The $(0,0)$ Fourier component of the flow represents a mean spanwise drift, while the $(k_x,0)$ components correspond
to spanwise-uniform fluctuations. 
Corresponding to WCC, in  our RNL$_N$ simulations we filter out both
the spanwise mean $(0,0)$ component  and the spanwise uniform $(k_x,0)$  fluctuation components 
for all cases  with channel widths $L_z>2 \pi$.  
Although spanwise uniform fluctuation components are crucial for sustaining turbulence in narrower minimal channels,   
as  noted  by Hamilton et al. \cite{Hamilton-etal-1995},
turbulence  is sustained for extended periods in our channel at $R=600$ when $L_z > 2\pi$.


\section{The fixed-point S3T equilibrium at $R=600$ and $L_z/h=2.2 \pi$ and its significance}

We first focus on the channel with spanwise spatial period $L_z/h=2.2 \pi$ which
supports an S3T equilibrium state to which
all RNL$_N$ simulations with $N \geq 2$ converge. 
This equilibrium consists of a streamwises mean flow, $\U_{eq}(y,z)$, 
with two low-speed streaks and two high-speed streaks, as illustrated in Fig. \ref{fig:S3T} (top),
and a rank two fluctuations covariance, $\C$.
The equilibrium mean flow and the covariance 
possess shift-rotate symmetry described by the transformation $[u,v,w](x,y,z)\to [-u,-v,w](-x+L_x/2, -y, z+L_z/4)$.
It is not possible to invert the second cumulant, $\C$,  for
the structures that make up the fluctuation field.
However, the rank of the covariance matrix $\mathbf{C}$ reveals the number of 
independent structures  comprising the fluctuation field, and
its two orthogonal eigenvectors identify the active subspace of fluctuations sustaining the turbulence,
but not the individual  structures sustaining the turbulence. 
These structures are the neutral modes of the operator in 
Eq. \eqref{eq:eRNLp} with mean flow $\mathbf{U}_{eq}$, and 
are determined by eigenanalysis of 
that operator.  These two neutral modes are  shift-rotate symmetric counterpart structures with phase speeds $c_r=\pm 0.106$ and with the structure of oblique waves centered around
the upper and lower streaks, respectively, as shown in Figures \ref{fig:P2} and \ref{fig:P3}.  Starting with a set of $N\ge 2$ randomly initiated ensemble members, the S3T system converges to the fixed point with these neutral structures comprising the fluctuation field in each ensemble but with amplitudes differing due to differing initializations.  Although the coefficients of the two modes generally differ in each ensemble, in sum these coefficients produce the  correct  square amplitude of each mode and also satisfy the non-interaction condition \eqref{eq:C}. 
Alternatively, once the neutral modes of $\U_{eq}$ have been determined by eigenanalysis, the S3T equilibrium can be realized most transparently by initializing using two members of the fluctuation ensemble, with the first member of the ensemble initialized by the fluctuation field of one of the neutral modes
and the second ensemble by the other neutral mode at the appropriate amplitude.

As we have noted, the neutral modes are not orthogonal in the energy metric and 
can interact with each other producing Reynolds stress divergence at their beat frequency revealed by exchange of energy with the mean flow (first cumulant).  Due to the interaction of the two neutral modes, 
the S3T fixed point can not be perfectly realized in an  RNL$_1$ simulation or its associated DNS. 
However, although the S3T equilibrium does not converge to a fixed point in an 
RNL$_1$ simulation, the resulting flow remains close to the fixed point of the S3T equilibrium and this proximity persists in DNS as revealed by the streaks that emerge in the DNS being persistent and reflecting the structure of the S3T equilibrium. 
Figure \ref{fig:P0} demonstrates the persistence of the streaks 
in DNS, while Figure \ref{fig:S3T} illustrates their structural 
similarity to the S3T fixed point.  Taken together these results confirm that the dynamics identified in ideal form by S3T fixed point equilibria underlie the turbulence seen in DNS and that the fluctuations responsible for sustaining the  turbulence in DNS are the same small number of structures comprising the active subspace identified by S3T.

\begin{figure}
    \centering
    \includegraphics[width=1.0\linewidth]{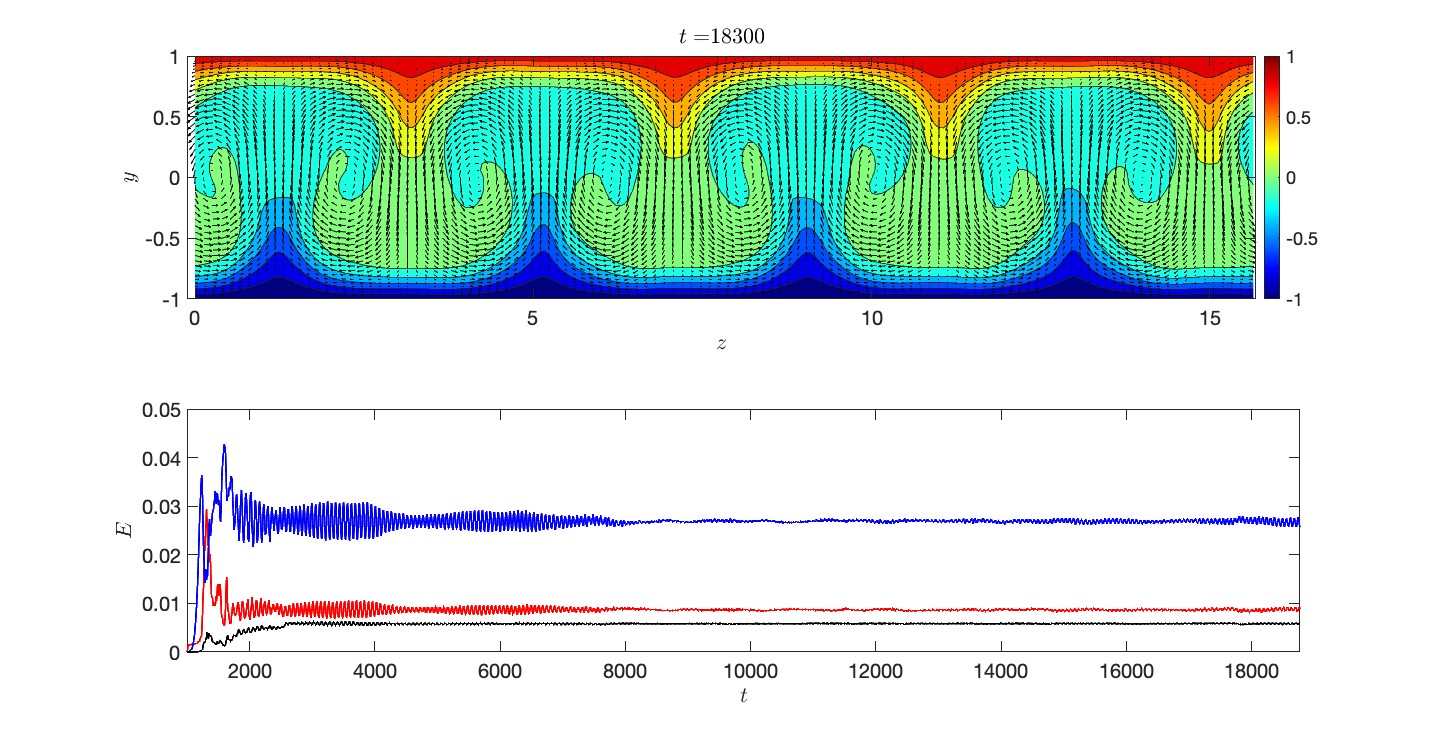}
 \caption{Top: Snapshot at $t=18300$ of the streamwise mean streak, $U_s$ (color levels),  and 
 vectors of $(W,V)$  from an RNL$_8$ simulation of WCC turbulence. The streamwise mean flow is almost an S3T equilibrium.%
  Bottom: Time evolution of the energy of the streak component (blue), of the fluctuations (red), and of the roll component (black) for the same simulation. The simulation was initiated from the laminar state at $t =1000$.
The flow was initially driven to turbulence using stochastic forcing applied until $t = 2500$.  The flow continued in a spatially and temporally turbulent state until it got arrested 
at $t \approx 3000$ to its  S3T attractor  state shown in the top panel.  
 The channel width is  $L_z=5\pi$ and the Reynolds number is $R=600$.}
    \label{fig:Lee1}
\end{figure} 

\begin{figure}
    \centering
\includegraphics[width=1.0\linewidth]{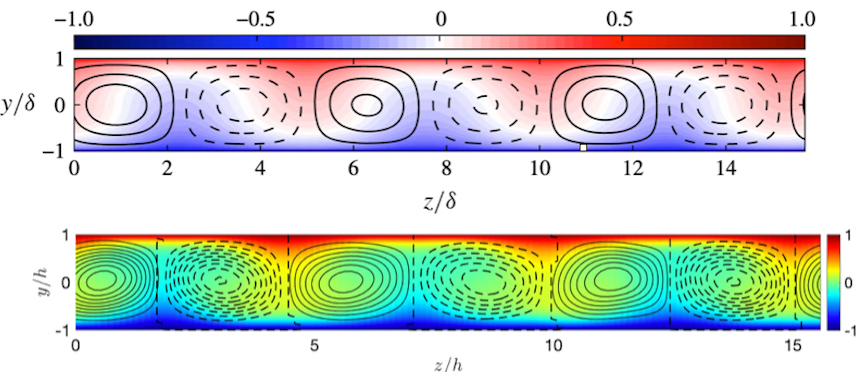}    
    \caption{Top: Time-averaged streamwise velocity $U(y,z)$ and  contours of the 
    time-averaged stream function
     of the $(V,W)$  obtained by Lee \& Moser \citep{Lee-Moser-2018}  in a DNS  of WCC at $R_\tau=500$ 
     in a channel with $L=5\pi$, $L_x=20 \pi$. 
 Bottom: The corresponding time-averaged streamwise-mean flow obtained in S3T simulations of WCC in a channel with $L_z=5\pi$ and $L_x=1.75 \pi$ at $R=600$. In this figure $y$ and $z$ are dimensional.
 }
    \label{fig:Lee2}
\end{figure}

The S3T equilibrium fixed point solution for WCC turbulence
is crucially dependent 
on the advection of the fluctuation component
of the flow by the roll velocities $V$ and $W$. Upon
switching off the $V$ and $W$ advection in the fluctuation
equations \eqref{eq:eRNLp} the fluctuations, which have been advected towards the center of the channel  by the streamwise rolls, relax immediately
towards the boundaries and 
the equilibrium state collapses.
The turbulence that ensues
is characterized by chaotically moving roll-streak structures. It is important to note that  the neutral modes are sustained 
principally by transfer of energy from the spanwise shear of the streamwise mean flow, $U(y,z)$,  
and that energy is transferred from these 
neutral modes to the roll component of the flow to maintain the fixed point WCC turbulence.
This  indicates that the S3T equilibrium is supported by a self sustaining process (SSP) 
\cite{Waleffe-2003}.   However, unlike the
unstable ECS equilibria, 
this S3T equilbrium is stable and the neutral modes sustaining this equilibrium
depend on the $V$ and $W$ velocities and are not inflection modes.  It is crucial that these modes are $V$ and $W$ velocity dominated given that the fixed-point solutions in WCC persist at high Reynolds number where inflectional modes would become singular, which is at variance with observation of continuously varying RSS characterizing WCC turbulence at high Reynolds number.  We find that the modes are dependent for their continuous structure on the $V$ and $W$ velocities distributing the fluctuating energy transferred from the mean streak throughout the flow.

The S3T equilibria persist in channels with widths near $L_z/h=2.2 \pi$. 
However, as the channel width deviates from this optimal value, the roll-streak structures are no loger accommodated  by the spanwise wavenumbers allowed and the equilibria are frustrated and undergo a sequence of bifurcations.  As the spanwise width changes the system transitions first to period 1 states (observed, for example, at $L_z/h = 1.225$), followed by a period doubling bifurcation to period 2  states (occurring, for example, at $L_z/h = 1.25$), and eventually to a chaotic regime. 
However, the chaos that emerges in the frustrated states is predominantly temporal in character: the roll-streak structures themselves remain spatially coherent and persistent, oscillating erratically with small amplitude about fixed spatial locations. Consequently, the time-averaged flow field reproduces roll-streak structures qualitatively similar to those of the S3T equilibrium at $L_z/h=2.2\pi$.

A representative example is presented in Figure \ref{fig:Lee1}, which shows a snapshot from an RNL$_8$ simulation 
of a channel with $L_z/h = 5 \pi$ and the temporal evolution of the 
energies of the streak, roll and fluctuation components of the flow from the laminar initial state. 
The fluctuations of the roll-streak structure about this time-averaged state are small.
Qualitatively, this S3T turbulent state reproduces with remarkable fidelity the large-scale roll-streak 
structure obtained by Lee \& Moser \cite{Lee-Moser-2018} in a DNS of a channel with identical spanwise width but different streamwise extent
and at a  substantially higher Reynolds number, $R_\tau=500$. The  streaks of the S3T  state in Figure \ref{fig:Lee1} equilibrate  at 
spanwise wavenumber 4, although  in other simulations in the same channel the flow settled to an S3T 
state with spanwise wavenumber 3 and with the time mean flow shown in
 \ref{fig:Lee2} (bottom), indicating
that this channel admits S3T equilibria with multiple spanwise wavenumbers. 
Correspondence  between the quasi-equilibria states seen in DNS simulations and those obtained using   S3T  indicates
that the large-scale structure in the high-Reynolds-number DNS 
is fundamentally supported by  the same fluctuation modes identified in S3T at lower Reynolds number.

\begin{figure}
    \centering
\includegraphics[width=1.0\linewidth]{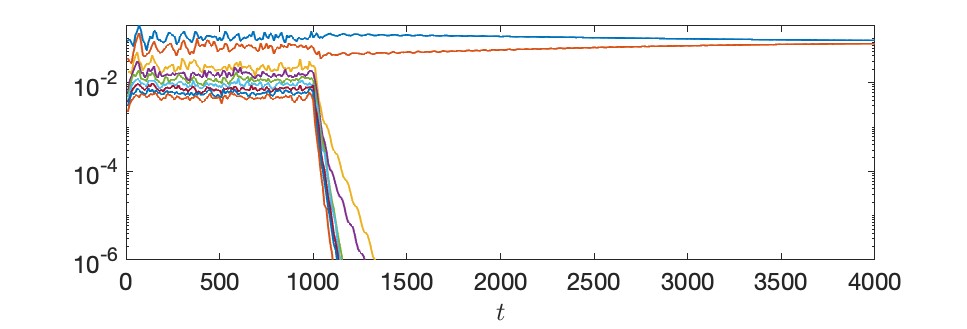}    
    \caption{Time series of the nine eigenvalues of the fluctuation covariance, $\C$, of an RNL$_9$ simulation as it converges to the S3T  
    equilibrium state. The number of non-zero 
eigenvalues of the covariance indicates the rank of the covariance. 
The simulation is stochastically excited for $t  <1000$ and during this period the rank of the covariance is equal to the
number of ensemble members, but two structures are dominant. 
When the stochastic excitation is switched off only the two dominant structures are sustained.
The subspace spanned by these
two structures is the active subspace of the turbulence at this Reynolds number. The 9 ensemble RNL eventually 
converges to the  equilibrium state that was also obtained in the RNL$_2$ and  shown in Fig. \ref{fig:S3T}.
 For a channel with  $z=2.2 \pi$ at $R=600$.    
 }
\label{fig:AP2}
\end{figure}

The S3T simulation in this channel  asymptotes  to the statistically stationary state  
depicted in Figure \ref{fig:Lee1}. 
During the initial transient phase spanning times $2000 < t  < 3000$ 
the statistical state dynamics  adjusts the coefficients of the 
active modes across the various ensemble components to eliminate their cross-interactions. 
This adjustment phase is observed in all simulations 
(see, for example, Figure \ref{fig:AP2}). Following this adjustment period, the flow either 
converges to a fixed point, as shown in Figure \ref{fig:AP1}, or settles into a spatially coherent 
state with mild temporal modulation, as shown in  Figure \ref{fig:Lee1}.
Even when the temporal dynamics exhibit chaotic behavior, 
the modulation is dominated by oscillations with a characteristic time scale of $O(70)$,
which we will show corresponds to the temporal signature of the self sustaining process (SSP). 
This  characteristic oscillation period appears during the approach to 
equilibrium shown in Figure \ref{fig:AP1} and throughout all simulations. This observation 
suggests that the time-dependent state  in Figure \ref{fig:Lee1} 
represents a weakly perturbed S3T equilibrium, in which the roll-streak structure 
instead of being at  equilibrium with its 
neutral modes is in an oscillatory finite amplitude nearby state, with principal period the natural period of oscillation 
of the roll-streak structure.

\section{The stability of the WCC S3T equilibrium}

The stability properties of the S3T equilibium 
could be found by eigenanalysis of the perturbation equations of the S3T system about this equilibrium
as was done previously \cite{Farrell-Ioannou-2012,Farrell-Ioannou-2024-S3T}.
Alternatively, in this work the decay rate, the oscillation frequency and the structure of the 
least damped mode of this S3T equilibrium state are determined from the 
approach of the RNL$_N$ simulations to the S3T equilibrium. 

\begin{figure}
\centering
\includegraphics[width=1.0\linewidth]{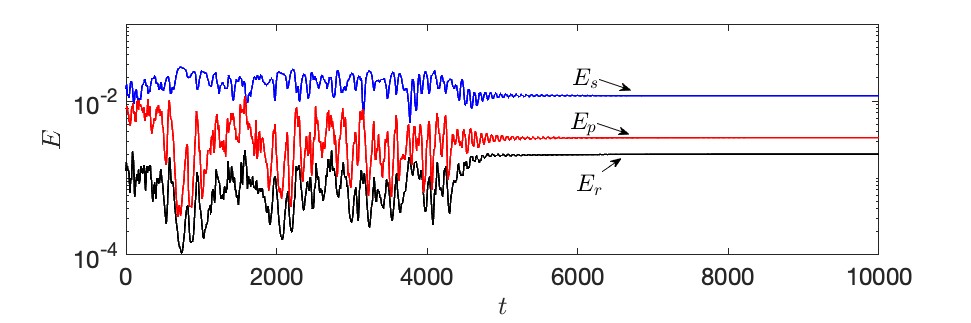}    
    \caption{Convergence to the S3T equilibrium state shown in Figure \ref{fig:S3T} by an RNL$_2$ simulation initiated  from a turbulent state which was  obtained in DNS.   Shown are the streak
energy density,  $E_s=[U_s^2]_{y,z}/2$ in blue;  the fluctuation energy density,
$E_p=[|\u'|^2]_{x,y,z}/2$ in red; and in black the roll energy density, $E_r=[V^2+W^2]_{y,z}/2$. 
For a channel with  $z=2.2 \pi$ at $R=600$ }
    \label{fig:AP1}
\end{figure}

\begin{figure}
    \centering
    \includegraphics[width=0.8\linewidth]{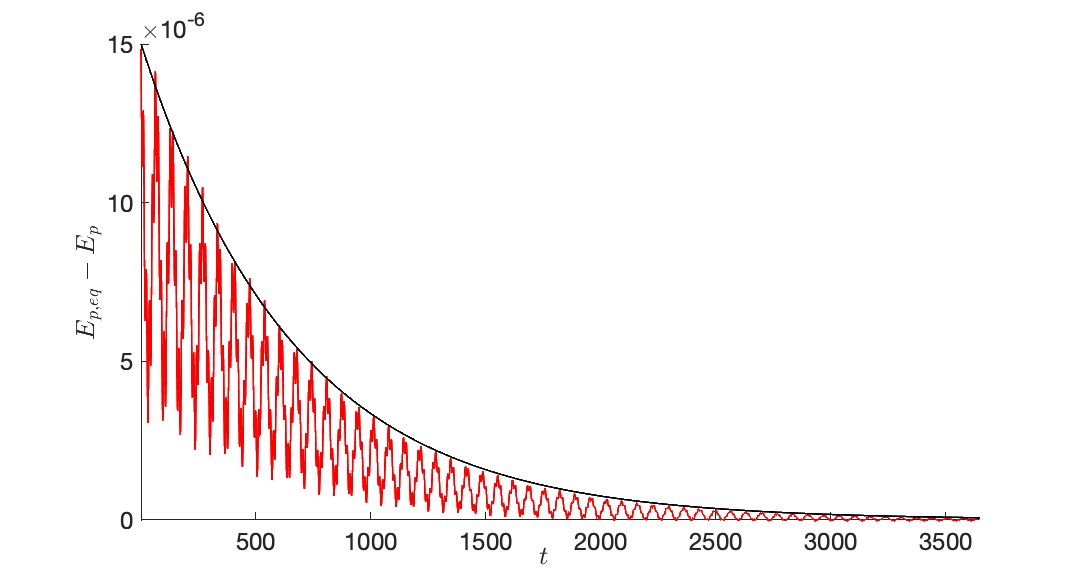}    
 \caption{Energy of the fluctuations, $E_p$, during the final approach to S3T equilibrium in an RNL$_2$ simulation. 
 This figure is a segment of the evolution of $E_p$ in the RNL$_2$ simulation shown in Fig. \ref{fig:AP1}.  The black 
 curve indicates that the equilibrium is approached exponentially at the rate $\sigma \approx 0.0015$ and with
 the period of the principal oscillation of the SSP which is $\approx 70$.  }
    \label{fig:Efinal}
\end{figure}

\begin{figure}
    \centering
    \includegraphics[width=0.8\linewidth]{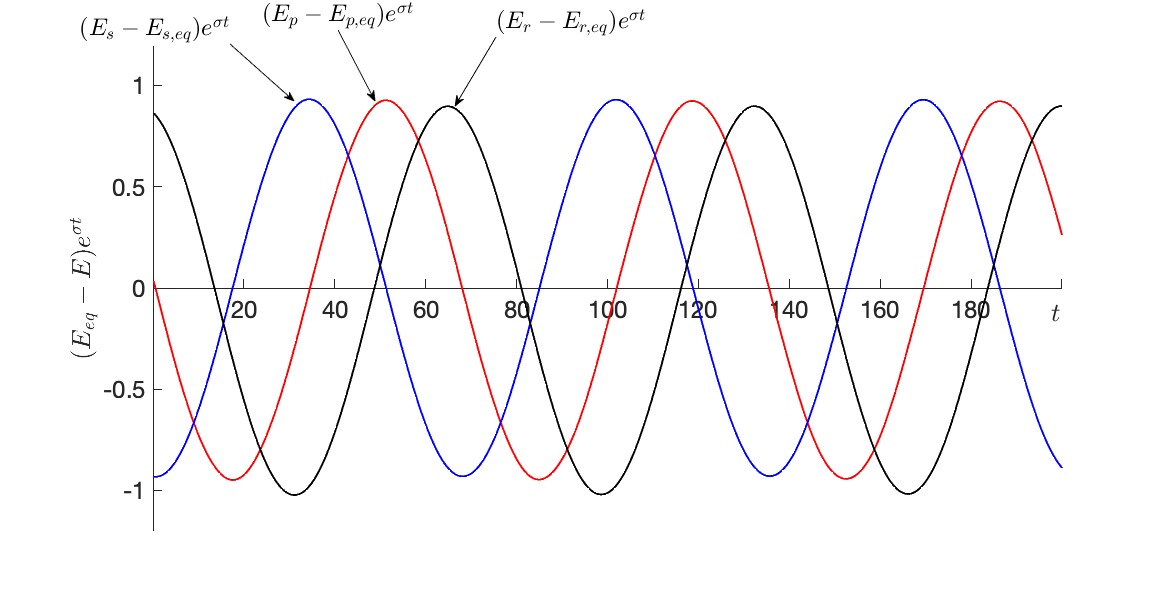}    
 \caption{Normalized streak, $(E_s-E_{s,eq}) e^{\sigma t}$ (blue),  energy of the fluctuations, $(E_p-E_{p,eq})e^{\sigma t}$ (red), and roll velocity, $(E_r-E_{r,eq})e^{\sigma t}$ (black),  as the equilibrium is approached in an RNL$_9$ simulation. All components
 of the flow approach equilibrium
 at the same rate $\sigma$.  This approach  rate is the same as that obtained in the  RNL$_2$ simulation  
 shown in \ref{fig:Efinal}. The oscillation period is also $\approx 70 $, as in  the RNL$_2$ simulation.
 Note that for small departures from the equilibrium $E_s-E_{s,eq}$ is   linear in $U_s-U_{s,eq}$, $E_r - E_{r,eq}$ is linear in
 $V-V_{eq}$ and $W-W_{eq}$, while $E_p-E_{p,eq}$ is quadratic in  deviations of the fluctuation velocity from the equilibrium value. 
 This figure shows that the streak amplitude, the fluctuation energy, and the amplitude of the roll components exhibit the characteristic period of the SSP. }
    \label{fig:Eallfinal}
\end{figure}

We have verified that the approach to the equilibrium state is an inherent property of S3T, independent of the RNL$_N$ simulation.
This allows insight to be gained into the mechanism of equilibration 
in the S3T system (cf. Figures \ref{fig:Lee1} and \ref{fig:AP1}).
For example Figure \ref{fig:AP1} shows the approach to the WCC equilibrium
in an RNL$_2$ simulation. In the first $5000$ time units of the simulation, the regulator modifies the modal mixture within the two ensembles establishing equivalent orthogonality thereby suppressing their interaction in the ensemble mean, which 
reflects a property of the specific RNL$_2$ simulation.
In contrast, the subsequent asymptotic approach toward the S3T equilibrium is the same
in all RNL$_N$ simulations and thus constitutes a universal S3T property.
In Figure \ref{fig:Efinal} we have redrawn 
the energy of
the fluctuation component,  $E_p$, 
as the equilibrium shown in  Figure \ref{fig:AP1} is approached. 
From this figure we obtain  
that  the decay rate of the least damped mode 
is about $\sigma = 0.0015$ and the period 
of this S3T mode is approximately  $70$.
We have confirmed  that the streak amplitude and the amplitude of the roll velocities 
decay to  their equilbrium values at the same rate as the energy of the  fluctuations.
This is a property of S3T stability that has as its variables the streamwise mean flow (the first cumulant)
and the quadratic covariance of the fluctuations. As a result when an S3T equilibriun is approached
the decay rate of the amplitudes of the streamwise mean flow towards its equilibrium value
will be the same as the decay of the quadratic fluctuation variables towards their equilibrium value.
In Fig. \ref{fig:Eallfinal} we
show the decay towards the equilibrium in
an RNL$_9$ simulation. In this figure
we have multiplied the fluctuations in the streak, roll velocity components and the 
fluctuations of the energy of the 
fluctuations with the decay rate $\sigma= 0.0015$
and normalized their amplitude in order to display  clearly the
common oscillation period of the three components of the flow and 
demonstrate that the  streak, the fluctuations and the roll component of the flow share
the same oscillation period, which is that of the SSP.

The same type of SSP oscillation with approximately the same period is observed in chaotic 
S3T states, for example for times $t > 10000$ in the $L=5 \pi$ channel
shown in Figure \ref{fig:Lee1}.
In all cases
the flow is dominated by two structures, as for example is evident 
in Figure \ref{fig:AP2},  and at all times 
the  SSP  is sustained  by interaction of the
roll-streak structure with these two dominant fluctuation modes, which were identified in our S3T analysis.

\section{Conclusions}
  The simplest manifestation of turbulence in wall-bounded shear flow is WCC turbulence, which exhibits clear fixed-point structure.  Failure to build a theory for turbulence in shear flow by first obtaining a theoretical understanding of the fixed point underlying WCC turbulence can be traced to failure to identify the analytic structure of this fixed point.  This failure in turn resulted from the tradition of analyzing shear flow turbulence dynamics using the NSE expressed in velocity state variables, in which the  fixed point underlying WCC turbulence lacks analytic expression.  However, when the NSE is formulated as a SSD, the fixed point solution underlying WCC turbulence is obtained immediately as the finite amplitude equilibrium proceeding from a linear modal instability of the SSD which provides an analytical solution for WCC turbulence as well as an analytical transition mechanism.  Once in possession of this solution, the entire dynamics underlying the transition to and maintenance of WCC turbulence becomes available for analysis, including the structure and physical mechanism underlying the eigenmodes supporting the turbulent state. 

This study uses the S3T formulation of the NSE SSD to demonstrate that  WCC turbulence is organized around a robustly attracting equilibrium fixed point SSD state. 
The components of this fixed point state comprise three analytically identified elements: the streamwise-averaged mean flow (first cumulant) and an associated pair of neutral eigenmodes  supporting the second cumulant.  The turbulence is therefore rank three with all three structures being analytically characterized in the S3T SSD framework.  Moreover, it is important to note that the limit cycle and the chaotic turbulent states are also essentially rank three, being supported primarily by the first cumulant and two fluctuation structures with zero Lyapunov/ Floquet exponent.

The rank three dynamics of the fixed point of WCC turbulence constitutes the entire turbulent state in the S3T SSD.  In the NSE there is in addition a complement of streamwise wavenumbers supported by nonlinear scattering that are essentially parasitic on the S3T state and do not appreciably influence the dynamics.  The structure of this complement can be studied by including a stochastic parameterization white in space and time  in the S3T closure to account for this nonlinear scattering  \citep{Nikolaidis-Ioannou-2022}.  
At high Reynolds numbers,  turbulent boundary layers form at the cross-stream boundaries,  but these do not appreciably influence the turbulence beyond their immediate vicinity \cite{Jimenez-Pinelli-1999}. 

The fixed point turbulence of WCC identifies the most simple manifestation of the SSP as comprising a RSS (first cumulant) supported by two neutrally stable modes of the first cumulant.  These modes are not traditional inflectional modes as their dynamics is contingent on advection by the roll component of the RSS, without which they do not exist.  The fixed point turbulence requires the existence of these modes and of the RSS that mutually supports them and is supported by them.  Together these structures comprise the S3T fixed point turbulent state.  A condition for the existence of the S3T fixed point turbulence for otherwise given parameters is a quantization condition on the spanwise wavenumber of the S3T RSS.  In the case of a spanwise constrained channel,  
fixed points exist over an interval of spanwise channel width, while outside this interval the fixed point S3T turbulent state 
settles to either a limit cycle or to a chaotic state.  
We refer to these non-fixed point states as frustrated states because the S3T system always needs to search for a 
fixed point for some period of time, if initiated other than at a fixed point, and adopts 
chaotic or limit cycle behavior as its search for the fixed point in phase space proceeds. 
When no fixed point exists due to failure by the channel to allow satisfying the quantization condition  
for existence of these neutral modes supporting it the chaotic search never ends and so the search becomes the statistical state of the turbulence. 
These states of turbulence are commonly observed in DNS (cf. \cite{Hamilton-etal-1995}).  In the limit of a wide enough 
channel the quantization condition becomes moot, which explains why fixed-point turbulent 
states were only discovered when turbulence was explored in wide enough channels.  

In summary, the equilibria and modes of the S3T and the companion modes of the first cumulant  are the fundamental basis on which an analytic solution for turbulence in shear flow is constructed. This basis supports the simplest turbulence in shear flow - the fixed point turbulence in  WCC.  Limit cycle and chaotic turbulent states are constructed on the same fundamental rank three structure as the fundamental fixed point turbulence.  The elements in all cases being the rank one S3T RSS (first cumulant) and the covariance of the fluctuations, which is composed of two eigenmodes, Floquet modes, Lyapunov modes depending on the spanwise quantization allowing the organizing SSD eigenmode mode or not.

 \section{Acknowledgments}
 The authors wish to thank  Professor Myoungkyu Lee for insightful discussions. This work was supported in part by the European Research Council under the Caust grant ERC-AdG-101018287. 

 \bibliographystyle{jfm}

\end{document}